# Self-regulated growth of candidate topological superconducting parkerite by molecular beam epitaxy

Jason Lapano[1], Yun-Yi Pai[1], Alessandro R. Mazza[1], Jie Zhang[1], Tamara Isaacs-Smith[2], Patrick Gemperline[2], Lizhi Zhang[3], Haoxiang Li[1], Ho Nyung Lee[1], Gyula Eres[1], Mina Yoon[1], Ryan Comes[2], T. Zac Ward[1], Benjamin J. Lawrie[1], Michael A. McGuire[1], Robert G. Moore[1], Christopher T. Nelson[1], Andrew F. May[1], Matthew Brahlek[1*]

[1]Materials Science and Technology Division, Oak Ridge National Laboratory, Oak Ridge, TN, 37831, USA
[2]Department of Physics, Auburn University, Auburn, AL, 36849, USA
[3]Department of Physics and Astronomy, University of Tennessee, Knoxville, TN 37996, USA
Correspondence should be addressed to *brahlekm@ornl.gov

**Abstract**:

Ternary chalcogenides such as the parkerites and shandites are a broad class of materials exhibiting a rich diversity of transport and magnetic behavior as well as an array of topological phases including Weyl and Dirac nodes. However, they remain largely unexplored as high-quality epitaxial thin films. Here, we report the self-regulated growth of thin films of the strong spin-orbit coupled superconductor $Pd_3Bi_2Se_2$ on $SrTiO_3$ by molecular beam epitaxy. Films are found to grow in a self-regulated fashion, where, in excess Se, the temperature and relative flux ratio of Pd to Bi controls the formation of $Pd_3Bi_2Se_2$ due to the combined volatility of Bi, Se, and Bi-Se bonded phases. The resulting films are shown to be of high structural quality, the stoichiometry is independent of the Pd:Bi and Se flux ratio and exhibit a superconducting transition temperature of 800 mK and critical field of $17.7 \pm 0.5$ mT, as probed by transport as well as magnetometry. Understanding and navigating the growth of the chemically and structurally diverse classes of ternary chalcogenides opens a vast space for discovering new phenomena as well as enabling new applications.

The ongoing search for materials that exhibit exotic quantum phenomena has led to a surge of research in low-dimensional chalcogen and pnictogen based compounds and intermetallics[1]. These materials possess the promising ingredients for novel behavior including magnetism[2–9], topology[10–14], and superconductivity[15–22], and, as such, may possess desirable properties such as quantized anomalous Hall effect[23,24], Weyl and Dirac nodes[3,25–27], negative thermal expansion[28,29], topological superconductivity[30–32], and may be key to realizing room-temperature applications[33] of topological materials. One structural family of interest that has been sparsely explored is the parkerites with the formula unit $A_3B_2X_2$ $(B_2A_3X_2)$ ($A$=Ni, Rh, Pt, Pd; $B$= Bi; $X$= S, Se)[34–38]. Numerous compounds of these exhibit superconductivity, including $Ni_3Bi_2S_2$[39], $Ni_3Bi_2Se_2$[39], $Pd_3Bi_2Se_2$[35], and $Rh_3Bi_2Se_2$[34]. Interestingly, all these possess a lower symmetry monoclinic structure, which opposes one of Matthias' rules for superconductivity[40,41]. $Rh_3Bi_2Se_2$ includes a charge density wave occurring below 250 K, which coexists with the superconducting state. Further, $Pd_3Bi_2S_2$ is a semimetal that is predicted to host exotic nontrivial band structures[1], and recent bulk crystal and thin film experiments show evidence of topological surface states[42,43]. In the same family of materials are the shandites, which also possess the $A_3B_2X_2$ stoichiometry ($A$ = Ni, Co, Rh, Pd; $B$=Pb, In, Sn, Tl; $X$=S, Se)[38,44]. Many of these materials exhibit exotic transport phenomena, in part derived from the kagome network of $A$-site atoms arranged in the structure, most notably in the ferromagnetic Weyl semimetal $Co_3Sn_2S_2$[38,45,46]. Thus, this family of structurally and chemically similar materials exhibit a tantalizing interplay of topology, magnetism, and superconductivity, and could serve as an ideal playground to explore novel quantum heterostructures and interfacial phenomena.

Both the parkerites and shandites are subclasses of the half-antiperovskites, which themselves are vacancy derivatives of the simple perovskite structure[47,48]. The structural evolution from the perovskite to half-antiperovskite parkerites and shandites is outlined in Fig. 1. In a typical cubic $ABX_3$ perovskite, $A$ and $B$ are cations that sit at the unit cell corners and center, respectively, while $X$ is an anion that occupies the unit cell face centers forming a 3D connected octahedral network, which is shown in Fig. 1(a). Here $X$ is most commonly O, but, for example, can also be F, N, and S, and the bonding is mostly ionic in nature. The structure is stable across a wide range of cations and anions with different atomic radii and charge states. This chemical diversity is accommodated through displacements of the cations, as well as distortions and rotations of the oxygen octahedra, and is made possible by a large stability window geometrically captured by the Goldschmidt tolerance factor. As shown in Fig. 1(b), the antiperovskites are characterized by simply switching the $A$-site cation position with the anion, leading to a chemical formula $A_3BX$, for example in $Mn_3GaN$. Though there is no loss of symmetry between these two classes, and in fact the same structural distortions can be found in both, it is useful to classify them as different materials since perovskites are typically ionically bonded whereas the bonding in antiperovskites is a mixture of predominately metallic and covalent[49]. The half-antiperovskite structures are derived by leaving half the $A$-site atoms vacant, giving a formula of $A_{3/2}BX$, or equivalently $A_3B_2X_2$. Further, the nature of the ordering of the vacancies dictates if they fall into the shandites or parkerites, which can drastically affect their ground state properties. The shandite compound is made by ordering the $A$-site vacancies in every other plane in the (111) direction, leading to the creation of a well-separated network of Kagome planes that can drive exotic magnetic and electronic phenomena[3,38,50–53]. In the parkerites, the vacancies order in 2 ways, as shown in Fig. 1(e-d). The $A$-sites can be vacant along every other (110) plane, leading to a monoclinic 2D layered-like structure (space group 12), or with alternate $A$-sites vacant along each (110) plane, leading to a cubic interconnected 3D structure (space group 199)[36,44,54,55]. Interestingly, superconductivity has primarily been observed in the 2D-type monoclinic parkerites but not in the cubic parkerites, and only reported in the $Pd_3Pb_2Se_2$ shandite compound at high pressures[56]. This suggests, again, a strong interplay between structure, chemical composition, and quantum phenomena.

A key challenge towards understanding and ultimately utilizing these novel materials is growing high-quality thin films of ternary compounds by molecular beam epitaxy (MBE)[57,58]. In contrast, the growth of high-quality binary chalcogenide thin films has been relatively simple given the existence of stoichiometric growth windows. This technique, also known as the three-temperature method, is widely utilized in the III-V semiconductor industry, as well as enabling GaAs/Ga$_{1-x}$Al$_x$As heterostructures to achieve mobilities in excess of 40,000,000 cm$^2$V$^{-1}$s$^{-1}$ [59,60], and is also the case of many binary chalcogenide topological insulators, such as Bi$_2$Se$_3$ or Bi$_2$Te$_3$[57]. In principle, the high vapor pressure element (Se, or Te) is supplied in excess compared to the low vapor pressure element (Bi). The substrate is kept above the sublimation temperature of the high vapor pressure element such that excess material is desorbed rather than being incorporated into the film, leaving only perfect stoichiometric material behind[61–63]. Key challenges for extending this to the ternary systems is that a growth window may only exist for a single element, but could emerge for multiple elements due to reduced/enhanced volatility of bonded components[64]; an example of the latter is the hypothetical shandite Co$_3$Sn$_2$Se$_2$, where the production and sublimation of highly volatile Sn-Se on the growing surface results in Sn-free Co$_7$Se$_8$ films[57,65]. Here, we report the synthesis of thin films of the ternary chalcogenide parkerite material Pd$_3$Bi$_2$Se$_2$ by MBE. We show that by tuning the growth temperature and elemental flux ratios supplied to the growing surface a self-regulated growth window emerges which is controlled by the Pd flux due to the higher volatility of Bi and Se. This results in high-quality single phase Pd$_3$Bi$_2$Se$_2$, which exhibits a superconducting transition of 800 mK as confirmed by transport as well as magnetometry measurements. The information gained here will enable understanding and navigating the growth of this class of ternary chalcogenides and enable observations of new phases and new applications.

Samples were grown using a home-built MBE system operating at a base pressure lower than 5×10$^{-10}$ Torr. Pd, Bi and Se were all supplied via thermal effusion cells. The cell temperatures were adjusted before growth to supply the desired flux, which was calibrated by a quartz crystal microbalance. The Pd flux was kept at approximately 2.5×10$^{13}$ cm$^{-2}$s$^{-1}$, while the Bi flux was adjusted to change the Pd:Bi flux ratio and tune stoichiometry within the film. Se was supplied in excess of 20×10$^{13}$ cm$^{-2}$s$^{-1}$ to overcome the high volatility and prevent formation of defects related to Se vacancies. Samples were grown on both SrTiO$_3$ (001) and SrTiO$_3$ (110) substrates. X-ray diffraction measurements (XRD) were performed on a Malvern Panalytical X'Pert$^3$ with a 4-circle goniometer using Cu $k_{\alpha1}$ radiation. Normal state transport measurements were performed in the van der Pauw geometry using pressed indium contacts in a Quantum Design Physical Property Measurement System down to a base temperature of 2 K. Millikelvin transport was probed in an Oxford Triton cryogen-free dilution refrigerator using Stanford Research SR860 lock-in amplifiers. Magnetometry measurements were performed with Quantum Design MPMS3 with iQuantum He3 option. Rutherford backscattering spectroscopy (RBS) measurements were conducted at Auburn University using a 6HDS-2 tandem, National Electrostatics Corporation Pelletron, with 2 sources for ions, an alphatross (RF source for production of He+) and SNICS source (source of negative ions by cesium sputtering), using a He$^{2+}$ ion beam energy of 1.972 MeV, an incident angle $\alpha = 0°$, an exit angle $\beta = 10°$, and a scattering angle $\theta = 170°$. Fits to the experimental data were completed using the analysis software SIMNRA (simnra.com). First-principles density functional theory (DFT) calculations were performed using the Vienna ab initio simulation package (VASP)[66]; more details can be found in the Supplementary Materials. Scanning transmission electron microscopy (STEM) measurements were taken using a Nion ultraSTEM C3/C5 aberration corrected microscope operating at 100 keV. The sample was prepared using a FEI Nova dual beam focused ion beam (FIB) using a Ga$^+$ liquid ion source.

MBE growth of ternary chalcogenide thin films is expected to be highly sensitive to the growth parameters, particularly deposition temperature, as well as the relative flux ratio among all three elements. For the current case of Pd$_3$Bi$_2$Se$_2$ the relative volatility of Se is much larger than Bi which in turn is much larger than Pd. For example, the cell temperatures used, which reflect relative volatility, are of order 200 °C for Se, 600 °C for Bi and 1300 °C for Pd. Further, all Bi-Se compounds are also relatively volatile in comparison to Pd[67]. This can be seen since Bi-Se can be used as a congruently sublimating binary source where the temperature is of order 400-500 °C[67,68]. We conjectured that a stoichiometric growth window can be found when Se is supplied well in excess of both Bi and Pd, and Bi is supplied at or in excess of the stoichiometric ratio with Pd. Specifically, on the growing surface the Se and Bi will physically adsorb onto the surface, then either (i) bond directly into the Pd$_3$Bi$_2$Se$_2$ structure, (ii) form a Bi-Se phase, which can then be incorporated into the Pd$_3$Bi$_2$Se$_2$ structure or evaporate as a compound, or (iii) Bi and Se can directly evaporate off the growing surface. As such, excess Se (>10×Bi) is necessary to overcome the high rate of evaporation of Se and enable both the reactions to Bi-Se compounds as well as the Pd$_3$Bi$_2$Se$_2$ to go forward. This method of excess Se flux has been explored in great detail in binary selenide compounds, and is necessary for the formation of stoichiometric thin films with minimal Se vacancies[69–71]. Therefore, by fixing the Se flux at >10×Bi enables the growth mechanisms to be systematically elucidated by varying only two parameters: the growth temperature as well as the Bi flux relative to the Pd flux (i.e. Bi:Pd > 3:2), which we discuss next.

To explore the temperature dependence of the growth four films were grown at 200 °C, 300 °C, 400 °C, and 500 °C at roughly the stoichiometric Pd:Bi flux ratio of 3:2 on SrTiO$_3$ 001 to a thickness of 40 nm. The resulting XRD 2$\theta$-$\theta$ scans are shown in Fig. 2(a). Single phase Pd$_3$Bi$_2$Se$_2$ films could be achieved between 300-400°C, where the predominate set of peaks are the monoclinic 00$L$ planes (orthogonal to the monoclinic $a$-$b$ plane), which indicates the films preferentially nucleated with the 2D-like vacancy-ordered planes parallel to the substrate surface. The additional peaks that can be observed off the main 002 (29.41°) peak are due to the $220$ (29.89°), $22\bar{2}$ (29.99°) and $40\bar{2}$ (30.4°) orientations of the film, which are all structurally similar to the perovskite 110 surface but with different vacancy orderings. At 200 °C, films segregated into Bi$_2$Se$_3$, Pd$_3$Bi$_2$Se$_2$, and other unidentified phases, which indicates additional Se is absorbed due to insufficient thermal energy. At 500 °C, in contrast, films became insulating and optically transparent rather than reflective, and XRD 2$\theta$-$\theta$ scans revealed a large misorientation of the Pd$_3$Bi$_2$Se$_2$ film, as well as emergence of an unidentified peak due to formation of a secondary phase. This can in part be understood as the inability to form a conformal film of Pd. This is likely since 400-500 °C coincides with the so-called 3/8-rule for growth of a metal film. At or below 3/8 of the melting point of a metal a conformal film can be achieved, and above this temperature the material will coalesce into islands[72]. For the case of the ternary oxide PdCoO$_2$, this has been similarly found to delineate the growth of conformal films versus islands[73]. This hints that the reluctance for Pd to bond (i.e. Pd is a noble metal) may result in the dominance of this simple metal-like behavior, but requires future study.

Together, these data show that across the approximate temperature range of 300-400 °C the Pd$_3$Bi$_2$Se$_2$ phase is stabilized in excess Se with stoichiometric Bi. Using this information, a series of 40 nm thick Pd$_3$Bi$_2$Se$_2$ films were grown to map out the growth mechanisms in excess Bi to show if there exists a stoichiometric growth window. Further, the substrate orientation was changed to SrTiO$_3$ 110, which more closely matches faces of Pd$_3$Bi$_2$Se$_2$. This can be seen by comparing the unit cells for the perovskite 110 to that of the parkerite 001, as shown in Fig. 1(g). Here, it can be seen that the Bi sites overlay well with Sr sites for the 110 SrTiO$_3$ surface, whereas the match to SrTiO$_3$ 001 is not nearly as good. Experimentally, it was found the growth on SrTiO$_3$ 110 reduced misoriented nucleation. Moreover, as is commonly used to aid nucleation and promote conformal coverage at a heterointerface, an initial 2 nm layer was grown at a stoichiometric flux ratio of 3:2 at 300 °C[69,74] (these conditions were chosen since it is the lowest temperature where there were no Bi-rich phases found). The sample was then heated to 400°C, at which time the remaining 38 nm film was deposited with the desired Pd:Bi flux ratio. Films were grown over a range of Pd:Bi = 3:1.8 to 3:5, and the resulting XRD 2$\theta$-$\theta$ scans are shown in Fig. 2(b). Films grown at low Bi-flux again showed secondary phases and misoriented domains, similar to the film grown at 500 °C, which suggests that the previously grown film was Bi-deficient in addition to not being conformal. Increasing the Bi-flux to well above the stoichiometric ratio to 3:5 led to fully 001 oriented films. We hypothesize that the orientation control through Bi-flux maybe related to effects of surface energy and nucleation (see below), which are likely highly dependent on stoichiometry.

A zoom-in of the 2$\theta$-$\theta$ scan about the 002 Pd$_3$Bi$_2$Se$_2$ peak for the film grown at 3:3.5, shown in Fig. 2(c) reveals Laue-oscillations about the main peak, which indicates the films are near atomically flat. Further, the rocking curve scan shown in Fig. 2(d) exhibits a narrow full-width at half-max (FWHM) of 0.033° ~118 arcseconds, indicating a large coherence among the 00$L$ planes, and, therefore, a very high structural quality. The 00$L$ lattice spacing was extracted, which for all films was between 6.054 Å and 6.061 Å, varying by less than 0.5% and was relaxed close to the bulk parameter of 6.069 Å. Further, all films were textured with grain sizes on the order of >1 μm with a slight degree of in-plane twinning and rotations from the substrate, as can be seen in the atomic force microscopy (AFM) images (Supplementary Materials Fig. S1). Further, high-angle annular dark-field imaging scanning transmission electron microscopy (HAADF-STEM) measurements were performed, and the results are shown in Fig. 2(e) where there are several notable features. First, the wide scale image confirms the lateral scale of grains are much larger than the viewing window. Second the zoom-in confirms that vacancy-ordered planes lay perpendicular to the substrate surface (parallel to the 110 planes of SrTiO$_3$) as well as the high coherency among the 00$L$ planes. This can be seen by comparing the zoomed in image to the corresponding structural model and composite unit cell shown in Fig. 2(f). Altogether, this confirms that the films grown are of high structural quality, but do not directly address the films' stoichiometry.

RBS, a direct probe of film stoichiometry, was performed on 2 different samples, and the results are shown in Table 1. The first sample was grown at the stoichiometric flux ratio of 3:2 Pd:Bi, while the second was grown at the Pd:Bi flux at a ratio of 3:3.6. Despite this, both samples had nearly identical film stoichiometries of Pd$_{3(\pm 0.06)}$Bi$_{1.81(\pm 0.04)}$Se$_{2.16(\pm 0.11)}$ and Pd$_{3(\pm 0.06)}$Bi$_{1.84(\pm 0.04)}$Se$_{1.97(\pm 0.10)}$, respectively. This confirms that the excess Bi and Se are not incorporated into the film as impurity phases or defects, but rather are desorbed off the growing surface. Interestingly, the results suggest the films are slightly Bi deficient as increasing Bi flux does not change the composition (i.e. Pd$_3$Bi$_{2(1-\delta)}$Se$_2$, with $\delta \lesssim 10\%$). Although, it is not clear if this is due to vacancies on the Bi site, if Pd is occupying some of the vacant sites in the half-antiperovskite structure, or some other mechanism. Remarkably, these results when combined with XRD show that film orientation is highly dependent on the Pd:Bi flux ratio, while the stoichiometry is nearly independent of this ratio when sufficient Bi is supplied at the proper growth temperature.

Altogether, this information enabled the creation of a schematic phase diagram, shown in Fig. 3, which highlights the conditions needed for self-regulated growth by MBE. The phase diagram was plotted with the vertical axis being the Pd:Bi ratio (with increasing Bi and fixed Se) and the horizontal axis temperature. It was found that at and below 250 °C, Bi$_2$Se$_3$ and Pd containing compounds were the primary phases where the Bi$_2$Se$_3$ constituents increase with increasing Bi flux. For temperatures above 300-450 °C there is sufficient thermal energy to create the $Pd_3Bi_2Se_2$ phase via desorption of excess Bi and Se, which confirms that the Pd flux controls the growth. This adsorption-controlled growth window (shaded region) is delineated by the solid to dashed lines that trend upward with increasing temperature, which qualitatively capture the increased propensity for desorption at increased temperature (i.e. for the upper curve, at higher temperature the phase boundary between $Pd_3Bi_2Se_2$ and Bi-rich phases increases to higher Bi vapor pressure (effectively flux), and, for the lower curve, at higher temperature the phase boundary between $Pd_3Bi_2Se_2$ and Pd-rich phases similarly increases to higher Bi-vapor pressures). Within the growth-window, the films are extremely flat and of high structural quality. Above 450 ºC, Bi and Se volatilize out of the film faster than it could be incorporated, leading to insulating, nonstoichiometric material. Thus, the optimal deposition strategy using this diagram was found to be a thin stoichiometric buffer layer at 300 ºC, followed by deposition of the remainder of the film at 400 ºC in excess Bi.

The XRD results for the stoichiometric series point to a relation between nucleation kinetics and the amount of Bi supplied to the growth front. Specifically, films grown at a stoichiometric flux ratio are single-phased but possess multiple crystalline orientations, and single domain films are only achievable at the highest Bi fluxes. First-principles density functional theory (DFT) calculations were employed to give insight into this novel effect (see Supplementary Materials for calculation details). Surface energy calculations were performed for the 4 observed film orientations for each surface being either mixed-terminated or Pd-terminated, with the atomic supercell structures used in the calculation. This mimics the likely scenario where Bi-deficient/Pd-rich conditions may favor mixed terminated surfaces and drive nucleation of these other orientations of $Pd_3Bi_2Se_2$. The results are given in Fig. 4, where panels a-h show the surface structures in both plane-view (top) and in cross-section (bottom), as well as the relative surface energy for the Pd chemical potential, $\mu_{Pd}$, in panel i. It was found the mixed terminated 001 surface was the most stable for all investigated chemical potentials, while the Pd terminated 001 surface exhibited the highest energy. This is consistent with the bonding geometry of $Pd_3Bi_2Se_2$ where the mixed termination splits the structure at the layered-like boundary and the Pd termination has the split within a layered unit (breaking more chemical bonds). This is further consistent with the experimental results which show that the predominate orientation is always the 001, which indicates that the $Pd_3Bi_2Se_2$ prefers growing as a layered unit assembled parallel to the surface. Similarly, for the other surfaces the calculations show that a Pd-terminated surface is higher in energy than the same surface with mixed-termination. Furthermore, these are significantly lower in energy than a Pd-terminated 001 surface. To understand how these correlate to the experimental observations consider the growth in Pd-rich conditions where the lowest energy surface is the 001. The excess Pd is likely accommodated in some portion as a combination of Pd antisite defects (likely on the Bi site) or as Bi-vacancies. However, for Pd rich conditions the formation of a full 001 Pd-terminated surface is very unlikely. Thus, the DFT results suggests that it may be far more likely to nucleate alternate surfaces that are lower in energy, which correlates well with XRD analysis. These calculation and experimental observations point out an interesting observation where film orientation may be controlled not simply by choosing a specific substrate but through the overall growth conditions, and is an interesting open question going forward.

Electrical transport and magnetization measurements give information into intrinsic properties, such as the character of superconductivity and aspects of the electronic band structure, as well as extrinsic properties such as defects incorporated during the growth. The resistivity is shown in Fig. 5(a) for a sample grown with Pd:Bi = 3:2.5, which reveal metallic transport behavior down to 2 K where the resistivity decreases with decreasing temperature. Overall, the residual resistivity ratios (RRR, ratio of the resistance at 300 K and resistance at 2 K) for all films in this series was found to be around 20 with no clear relation to the growth conditions, as shown in the inset of Fig. 5(a). This shows that growth conditions are not the limiting factor for transport mean free path and points to an additional mechanism. Likely this could be either grain boundaries, dislocations, film thickness, or possibly the Bi deficiency. Regarding the latter, however, the percent-level of native Bi-deficient defects found in RBS would likely severely limit the overall mean free path to a few nanometers, and thereby the RRR. This will be further addressed next in the context of Hall effect measurements.

Magnetic field dependent Hall effect measurements can reveal additional insight into the electronic properties of these films. Hall effect was measured at a temperature of 2 K and magnetic field up to 14 T. The data was then anti-symmetrized to remove the unintentional mixing of the longitudinal component into the Hall component. The resulting data, shown in Fig. 5(b), revealed a dramatic non-linear character. This indicates intrinsic multicarrier behavior in the films due to both electron and hole conduction. This can be seen by a change in sign of the slope of the Hall effect in going from the low-field to the high-field limits. Data was then fit using a 2-carrier electron and hole model to determine the carrier type and mobility (see Supplementary Materials for more details). The dominate carrier type was found to be *p*-type (hole-like), with approximately $2-4\times10^{22}$ cm$^{-3}$ with mobility of ~100 cm$^2$V$^{-1}$s$^{-1}$, and a small electron pocket contributed between $1-3\times10^{19}$ cm$^{-3}$ and mobility of ~1000 cm$^2$V$^{-1}$s$^{-1}$. There was no discernable dependence on the growth conditions. Overall, the mixed character found in the Hall effect data is likely not anomalous. It is quite typical that good metals exhibit multiband transport due to band filling and subsequent zone-folding[75,76]. To observe such multicarrier behavior requires a large mobility, such that the inverse mobility is less than the maximum field scale used (i.e. $\mu^{-1} < B_{max}$). Here we observe the crossover to be of order several Tesla indicating a mobility of order of a thousand cm$^2$V$^{-1}$s$^{-1}$. This is consistent with the low-density higher mobility electron pocket dominating the response at low-field, which gives way to the high-field response being dominated by the lower mobility hole. From this mobility scale (several-hundred to thousand), the transport mean-free-path can be estimated from a simple free electron model where $l_{mfp} = \hbar\mu/e(3\pi^2 n_{3d})^{1/3}$, where $\hbar$ is Planck's reduced constant, $\mu$ is the mobility, $e$ is the electron charge, $n_{3D}$ is the carrier density. This yields a $l_{mfp} \approx$ 100-1000 nm, which is consistent with the scale of both the grains (~μm) as well as the thickness (depending on the level of specular vs non-specular scattering, surfaces can be the dominate scattering mechanism with $l_{mfp} \gg$ thickness)[77].

Lastly, polycrystalline bulk Pd$_3$Bi$_2$Se$_2$ samples have been reported to be a low temperature, type II s-wave superconductor[35]. Low-temperature AC-magnetic susceptibility magnetometry measurements were carried out on a stoichiometric film down to 300 mK for in-plane (IP) and out-of-plane (OOP) geometries, measured on cooling in zero applied DC field, with an AC field of 2.5 Oe and frequency of 75.7 Hz. DC SQUID measurements performed using the same instrument, with ZFC followed by FC cooling at each field. The results of the AC magnetometry measurements are shown in Fig. 6(a-b), respectively. Superconductivity can be seen to arise where there is a large drop in the in-phase component and a rise in the out-of-phase component, which indicates the onset of diamagnetic screening[78]. This occurs at approximately $T_C \approx$ 800 mK for both the IP and OOP orientations. The unusual temperature dependence observed in the OOP geometry can likely be attributed to the extreme demagnetization effects in this geometry[79]. Similarly, resistance measurements were performed down to 25 mK in an OOP orientation and the results are shown in Fig. 6(c). Here, the resistivity sharply drops into the zero-resistivity states at 800 mK, which agrees well with magnetization measurements. Further insight into the superconducting properties were gained by detailed temperature sweeps at different magnetic fields. In Fig. 6(c) the results are shown for the transport measurements. Here, with increasing magnetic field the transition into the superconducting phase is successively pushed to lower temperatures with fields >16 mT quenching the superconductivity down to 25 mK. Also, as the magnetic field was increased the width of the transition into the superconducting phase increased. Specifically, for zero-field the width is less than 10 mK, and above 14 mT, the width increased to about 100 mK consistent with weak-link behavior observed in textured thin films[80]. The field dependence of the superconductivity transition is captured by plotting $T_C$ vs $\mu_0 H$, shown in Fig. 6(d), where $T_C$ was taken at the onset and at half the normal state resistance in transport and from the onset in DC SQUID measurements. The slope of $\mu_0 H_C$ near $T_c$ was found to follow the Werthamer-Helfand-Hohenberg (WHH) theory for a type-II superconductor in the dirty limit[35,81]. The critical field at zero temperature ($H_{c2}(0)$) is extracted using the WHH theory for the slope of $\mu_0 H_{C2}(T)$ as $T \to T_C$:

$$\mu_0 H_{C2}(0) = -0.693 T_C \left.\frac{d(\mu_0 H_{C2}(T))}{dT}\right|_{T \to T_C}. \tag{1}$$

Here $\mu_0 H_{C2}(0)$ was extracted to be $17.7 \pm 0.5$ mT. From this the functional form for $T_C$ vs $H$, given by

$$H_{C2}(T) = H_{C2}(0)\left(1 - \left(\frac{T}{T_C}\right)^{1.44}\right), \tag{2}$$

is plotted as a dashed curve in Fig. 6(d), with $H_{C2}(0)$ given by the WHH theory. It can be seen that the data is in excellent agreement with this function in the zero-temperature limit. Further, the coherence length at zero temperature ($\xi_{GL}(0)$) is estimated to be around 140 nm from the following relation,

$$\mu_0 H_{c2}(0) = \frac{\phi_0}{2\pi \xi_{GL}(0)^2}, \tag{3}$$

where $\phi_0$ is the magnetic flux quanta. This value is significantly larger than previous experiments on bulk polycrystalline samples (32.3 nm)[35]. Additionally, this coincides with the order of magnitude of the transport mean-free-path estimated from the Hall effect observed in the film, which, again, suggests either grain boundaries or the surfaces are the limiting defects in these samples and points to a high crystalline quality within each grain.

In conclusion, we have demonstrated thin film synthesis of the strongly spin-orbit coupled superconductor $Pd_3Bi_2Se_2$ using molecular beam epitaxy. By carefully controlling growth conditions, we have shown the existence of a self-regulated growth window resulting in highly reproducible thin film growth with high structural quality. This is confirmed via XRD and RBS measurements. Films grown within this window exhibit metallic conductivity with relatively high residual resistivity ratios, superconductivity below 800 mK, and relatively large coherence length of ~140 nm. Synthesizing high-quality thin film materials is the first step in the scientific process and key questions remain for $Pd_3Bi_2Se_2$ and the parkerites and shandites, in general. Of particular importance is understanding and characterizing the topology of the band structure. This can be answered by a combination of first-principles calculations and photoemission spectroscopy, which also will give insight into the strong multiband behavior observed here. From a materials perspective, the role of defects is of critical importance. It remains unknown why $Pd_3Bi_2Se_2$ is found to be intrinsically Bi deficient and if this is the stable state or if this is controlled by the volatility of Bi and Se specific to MBE growth. Growth of high-quality bulk crystals will enable illumination of this question. Secondly, the observed stoichiometric-dependent nucleation suggests that this may be used to create films or nanostructures with different orientations that are independent of the substrates' surface structure. Lastly, $Pd_3Bi_2Se_2$ is found to nucleate better on the $SrTiO_3$ (110) surface compared to the (001) surface, due to the better surface match. However, there, the films are still found to be composed of in-plane grains, which may be the limiting factor for transport. Therefore, understanding what determines how the parkerites nucleate on these surfaces, their overall interfacial structure, as well as finding if there exists a better substrate, is of critical importance for future work, which will push towards higher film quality. Overall, this work opens the potential for thin film synthesis of the parkerites and shandite classes of materials, which when coupled with their chemical and structural compatibilities, make a promising playground to explore the interplay of spin-orbit coupling, topology, magnetism, and superconductivity.

## Data Availability

The data that support the findings of this study are available from the corresponding author upon reasonable request.

## Supplementary Materials

See Supplementary Materials for additional structural measurements, first principles calculation methods, as well as details regarding the multicarrier transport analysis. Supplementary Materials also contains references[66,82,83].

## Acknowledgements

This work was supported by the U. S. Department of Energy (DOE), Office of Science, Basic Energy Sciences (BES), Materials Sciences and Engineering Division (growth, structure, and electron microscopy), and the National Quantum Information Science Research Centers, Quantum Science Center (transport). This research used resources of the Oak Ridge Leadership Computing Facility and the National Energy Research Scientific Computing Center, DOE Office of Science User Facilities. RBS measurements at Auburn University was supported by Air Force Office of Scientific Research award number FA9550-20-1-0034.

# Figures

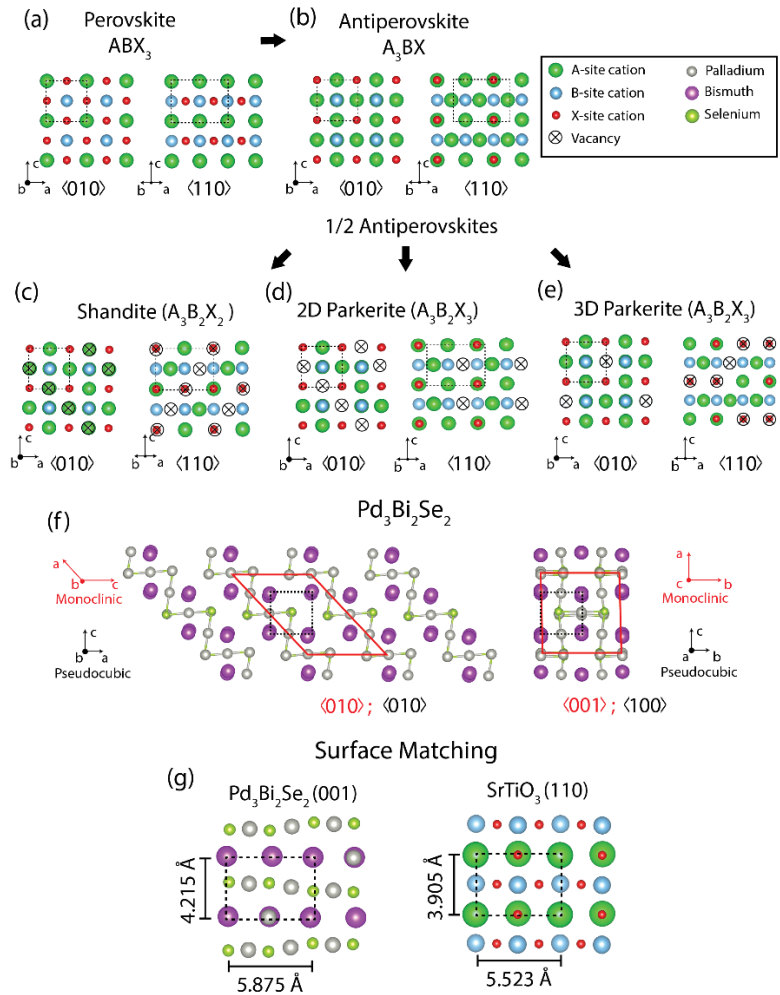

**Figure 1.** Structural evolution from the cubic perovskite to the 2D-parkerite $Pd_3Bi_2Se_2$. (a) The crystal structure of the perovskite. (b) Exchange of the *A*-site and *X*-site ions lead to the antiperovskite structure. (c-e) The half-antiperovskites are derived by vacant planes of the *A*-site along the pseudocubic (111) for shandites (c), or the (110) for parkerites (d, e). (f) The layered-like structures of the 2D-parkerite $Pd_3Bi_2Se_2$ in the proper monoclinic symmetry, with the corresponding monoclinic and pseudocubic unit cells highlighted. (g) The $Pd_3Bi_2Se_2$ (001) and $SrTiO_3$ (110) surfaces and lattice parameters are given to show the relative orientation and structural similarity between the film and substrate, in which bismuth aligns on strontium, selenium on titanium, and palladium on oxygen.

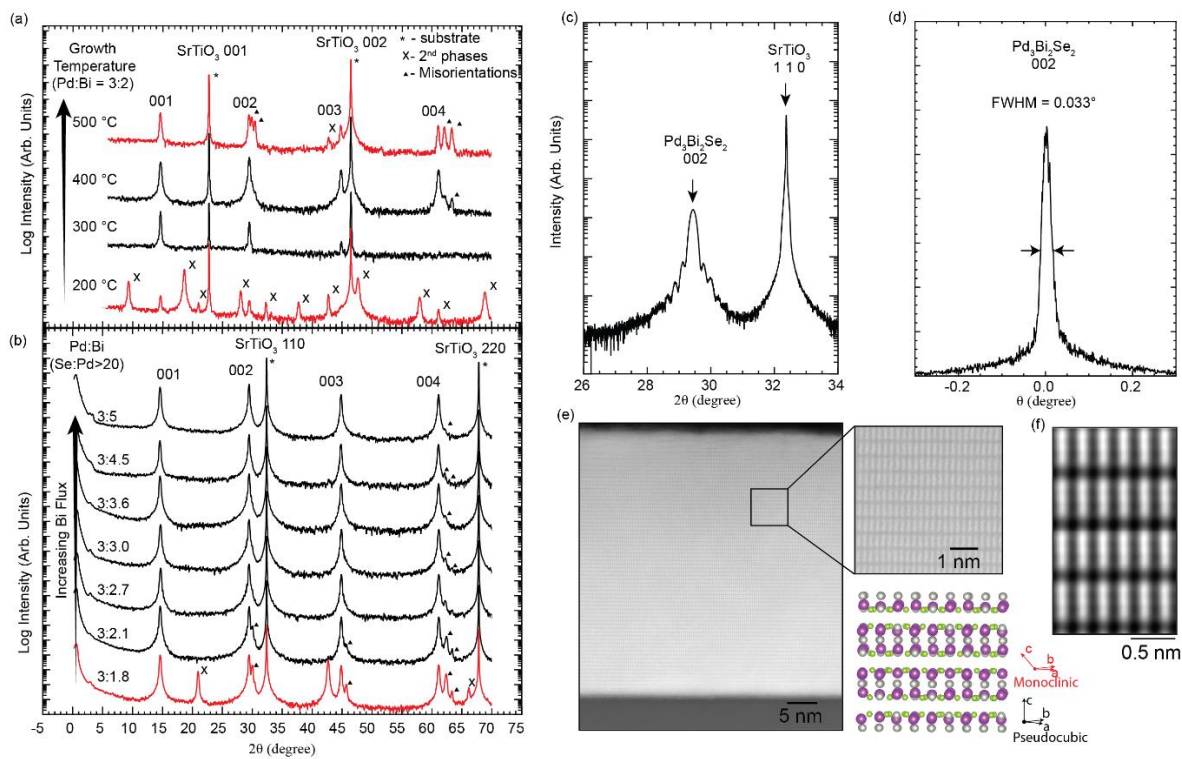

**Figure 2** (a-b) $2\theta$-$\theta$ XRD scans of the $Pd_3Bi_2Se_2$ grown at different temperatures and Pd:Bi fluxes are given in (a) and (b), respectively. Samples in (a) were grown on $SrTiO_3$ (001) at Pd:Bi=3:2, while all subsequent data were from films grown on $SrTiO_3$ (110) with flux ratio specified. Peaks from misorientations and secondary phases are marked by triangles and crosses, while the substrate peak is marked by an *. (c-d) Higher resolution scan of a film grown at the optimal conditions of 3:3.5 Pd:Bi exhibits Laue-oscillations and no second phase or misorientation (c), as well as a sharp rocking curve of 0.033° about the 002 peak. (e). HAADF-STEM image of $Pd_3Bi_2Se_2$ showing the 2D-like layer units, as seen in the corresponding structural model. (f) Composite unit cell obtained by breaking the image down into individual unit cells, averaging their intensity into one unit cell, which provides a higher resolution representation of the data, and an excellent agreement to the structural model.

**Table 1.** Rutherford backscattering spectroscopy results of $Pd_3Bi_2Se_2$ film stoichiometry compared to the flux supplied to the growing surface as measured in situ using quartz crystal microbalance. Films grown at nominal stoichiometric Pd:Bi flux ratio shows nearly identical film stoichiometry to one with excess Bi, exemplifying the self-regulated growth. Estimated error is roughly 2% for flux measurements, and for RBS results 2% for Pd and Bi and 5% for Se (see Supplementary Materials for RBS data and fitting).

|          | $Pd_3Bi_2Se_2$ Stoichiometric Growth | | $Pd_3Bi_2Se_2$ Excess Bismuth | |
|---|---|---|---|---|
| Elements | Normalized Flux | **Normalized RBS Ratio** | Normalized Flux | **Normalized RBS Ratio** |
| Palladium | 3±0.06 | **3±0.06** | 3±0.06 | **3±0.06** |
| Bismuth | 1.96±0.04 | **1.81±0.04** | 3.77±0.04 | **1.84±0.04** |
| Selenium | 23.1±1.2 | **2.16±0.11** | 22.2±1.1 | **1.97±0.10** |

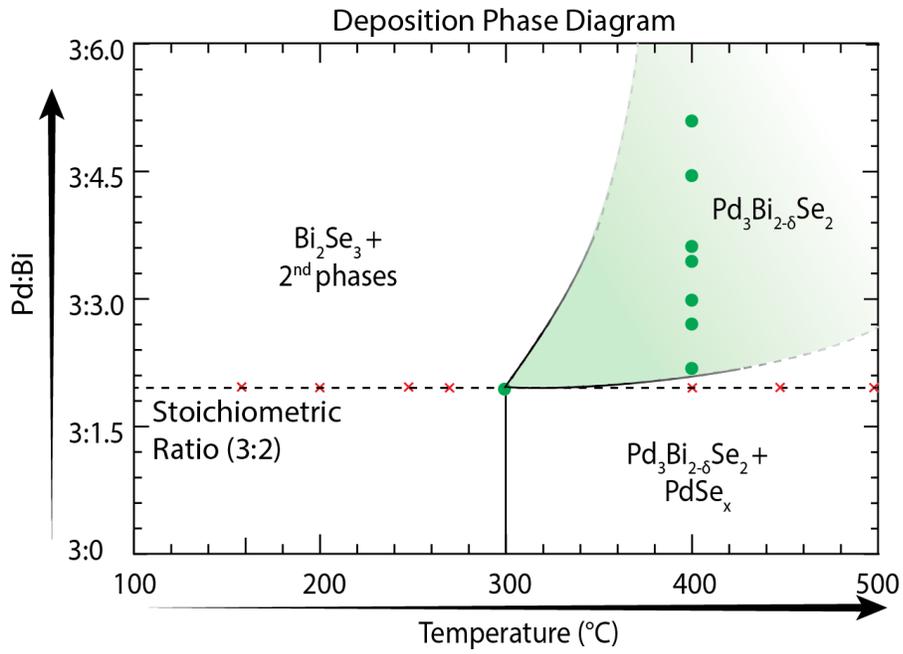

**Figure 3** Molecular beam epitaxy growth phase diagram for the ternary material $Pd_3Bi_2Se_2$ based on structural data in Fig. 2 and RBS measurements shown in Table 1. The shaded region highlights the conditions where high quality $Pd_3Bi_2Se_2$ can be grown in a self-regulated growth mode.

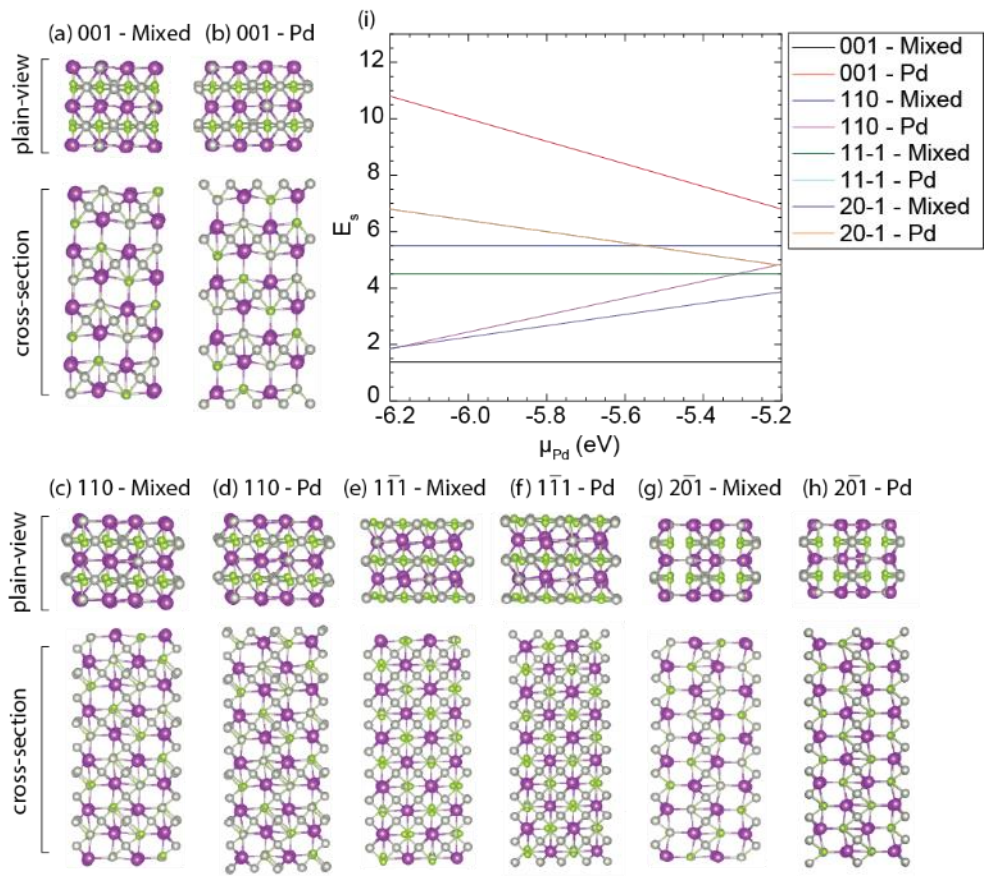

**Figure 4.** (a-h) Models of the surface structures, matching the various peaks observed in Fig. 2, with either mixed terminations (a, c, e, and g) or Pd terminations (b, d, f, and h), as indicated. The top row is the structures in plane-view and the bottom row is the structures in cross-section. (i) Relative energy of the various surfaces versus the Pd chemical potential, $\mu_{Pd}$.

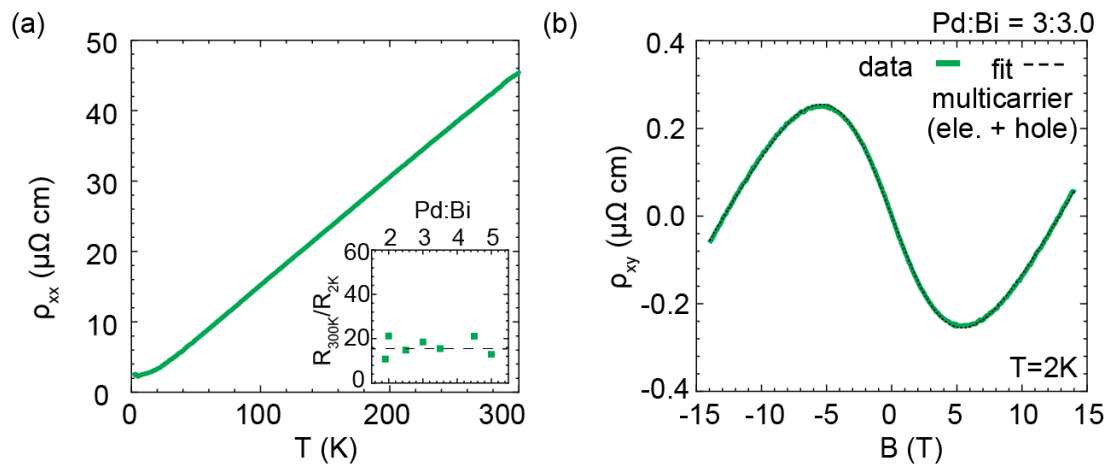

**Figure 5.** (a-b) Resistivity and Hall effect results for a sample grown at 400 °C in excess Bi (Pd:Bi ~3:3.0). The insets in (a) shows the residual resistivity ratio (RRR=$R_{300K}/R_{2K}$) versus flux ratio. In (b) the solid green line is the data, and the dashed black line is the fit to the multicarrier model (see Supplementary Materials).

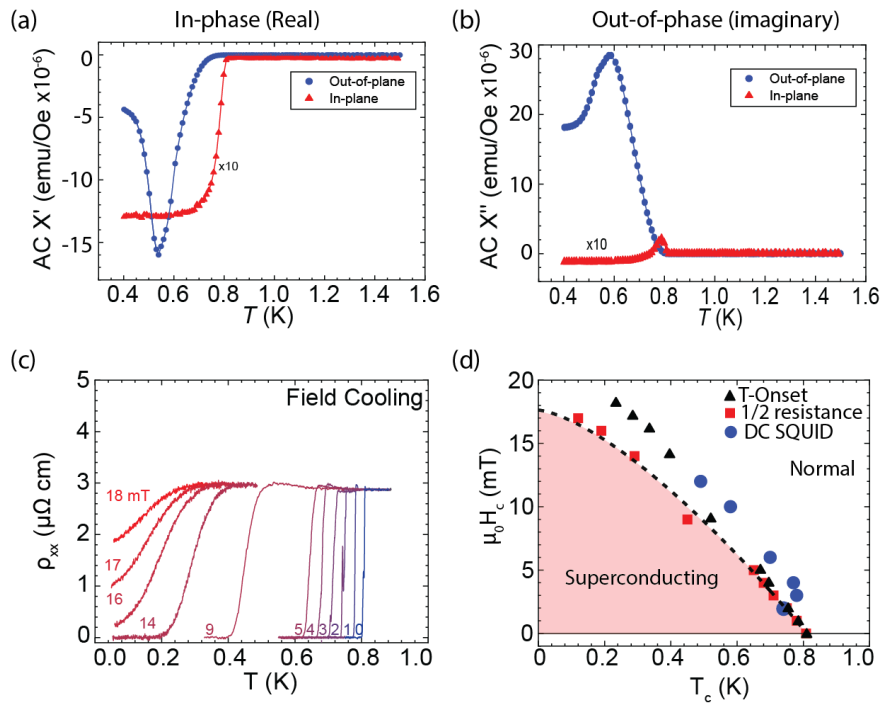

**Figure 6.** (a-c) Temperature dependence of superconductivity as probed by AC-magnetometry (a- b) in the in-plane and out-of-plane geometries, respectively, and the resistivity in the out-of-plane geometry (c). (d) Critical temperature versus critical magnetic field compiled from the data shown in (a-c) (symbols) and fit using the Werthamer-Helfand-Hohenberg theory (dashed line).

# Supplemental Materials
# Self-regulated growth of candidate topological superconducting parkerite by molecular beam epitaxy

Jason Lapano[1], Yun-Yi Pai[1], Alessandro R. Mazza[1], Jie Zhang[1], Tamara Isaacs-Smith[2], Patrick Gemperline[2], Lizhi Zhang[3], Haoxiang Li[1], Ho Nyung Lee[1], Gyula Eres[1], Mina Yoon[3], Ryan Comes[2], T. Zac Ward[1], Benjamin J. Lawrie[1], Michael A. McGuire[1], Robert G. Moore[1], Christopher T. Nelson[1], Andrew F. May[1], Matthew Brahlek[1*]

[1]Materials Science and Technology Division, Oak Ridge National Laboratory, Oak Ridge, TN, 37831, USA

[2]Department of Physics, Auburn University, Auburn, AL, 36849, USA

[3]Center for Nanophase Materials Sciences, Oak Ridge National Laboratory, Oak Ridge, TN, 37831, USA

[4]Department of Physics and Astronomy, University of Tennessee, Knoxville, TN 37996, USA

Correspondence should be addressed to *brahlekm@ornl.gov

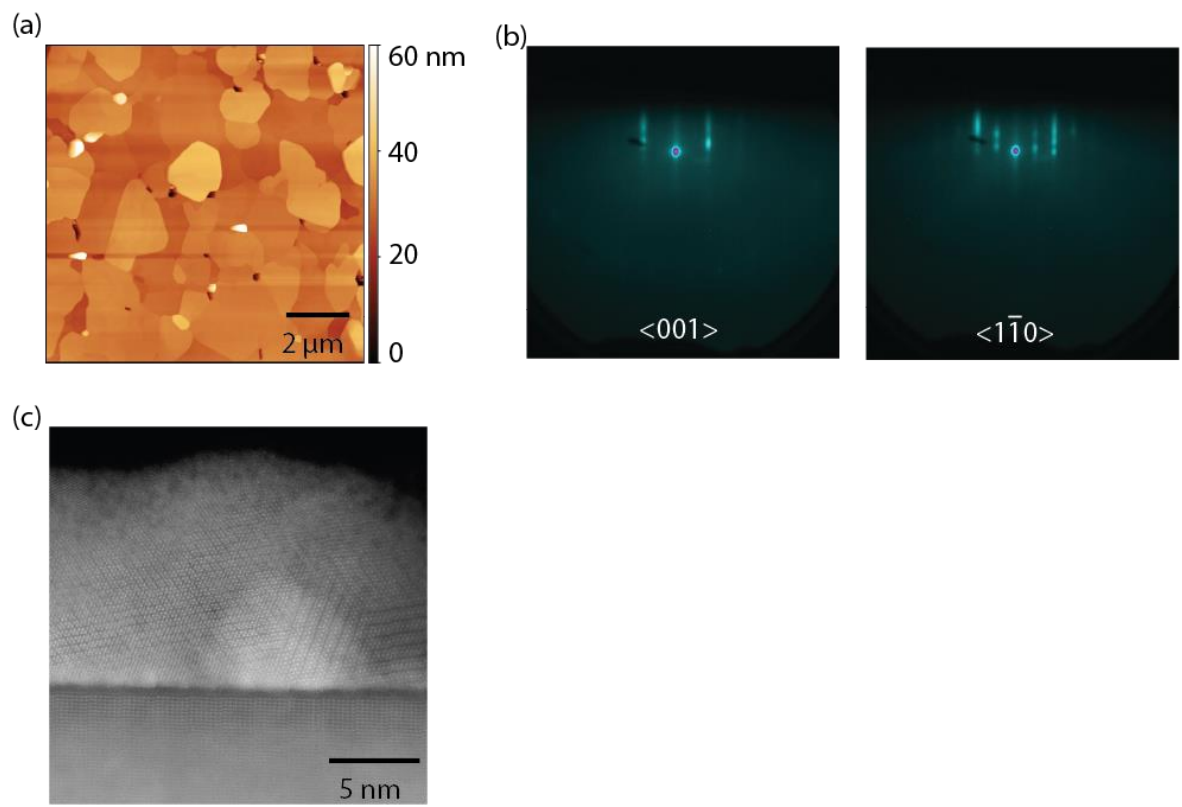

**Fig. S1** (a) Atomic force microscopy image of a representative $Pd_3Bi_2Se_2$ sample showing flat, fully connected grains. (b) Reflection high-energy electron diffraction (RHEED) images taken along the high symmetry directions of a $Pd_3Bi_2Se_2$ grown on $SrTiO_3$ (110) show streaky patterns indicative of a smooth growth front. (c) Small areas of misoriented grains can also be seen interrupting the larger 001 oriented grains in STEM.

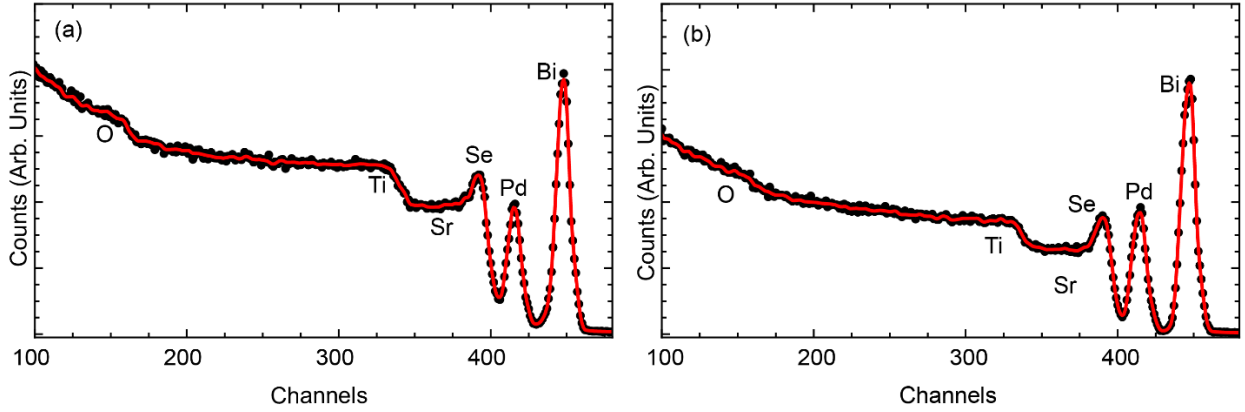

**Fig. S2** Rutherford backscattering spectroscopy data (symbols) along with fits (solid red line) for the two films shown in Table 1 of the main text. The sharper peaks emanated from the elements in the films (with increasing mass and channel #, Se, Pd, then Bi), while the longer plateau-like peaks are from the elements in the substrate (with increasing mass and channel #, O, Ti and Sr). (a) The film grown under stoichiometric conditions and (b) the film grown in excess Bi.

**First principles calculations**

All the calculations are based on first-principles density functional theory (DFT) using the Vienna ab initio simulation package (VASP)[1] with the projector augmented wave method; a generalized gradient approximation in the form of Perdew–Burke–Ernzerhof is adopted for the exchange-correlation functional[2], the energy cutoff of the plane-wave basis sets is 400 eV. We employ 3×4×6, 3×4×1, and 2×4×3 Monkhorst-Park k-point meshes for the unit cell of $Pd_3Bi_2Se_2$, high-symmetry surfaces, and 2×1×2 supercells, respectively. In all calculations, all atoms are fully relaxed until the residual forces on each atom are less than 0.01 eV/Å, and a 15 Å vacuum layer is used for the surface calculations. We evaluated the surface energies for the various facets of $Pd_3Bi_2Se_2$ shown in Figure 4 of the main text, defined as $E_s = (E_{tot} - mE_{ref} - \sum_i n_i \mu_i)/2$, where $E_{tot}$ is the total energy of a slab, $E_{ref}$ is the total energy of bulk $Pd_3Bi_2Se_2$, $m$ is the total number of unit cells in the slab model, $n_i$ is the number of $i$ atoms ($i$ = Pd, Se, and Bi), and $\mu_i$ is the chemical potential of each species. In thermodynamic equilibrium, $\mu_{Pd}$, $\mu_{Se}$ and $\mu_{Bi}$ satisfy the condition, $\mu_{Pd3Bi2Se2} = 3\mu_{Pd} + 2\mu_{Se} + 2\mu_{Bi}$, where $\mu_{Pd3Bi2Se2}$ is the chemical potential of $Pd_3Bi_2Se_2$. We choose the chemical potential of Pd in between that of bulk Pd and from the equilibrium condition to bulk $Pd_3Bi_2Se_2$ and bulk Se and Pd, i.e., $(E_{ref}-8*E_{Se}-8*E_{Bi})/12 \leq \mu_{Pd} \leq E_{Pd}$, where $E_{Se}$ = -3.5 eV is the total energy per Se atom of the bulk Se, $E_{Pd}$ = -5.18 eV is the total energy per Pd atom of the bulk Pd and $E_{Bi}$ = -3.89 eV is the total energy per Bi atom of the bulk Bi, then we have -6.06 eV $\leq \mu_{Pd} \leq$ -5.18 eV.

**Multiband transport model**

To fit the experimental Hall data a semiclassical multiband model was used. Here, individual transport channels have distinct mobilities ($\mu_i$) and sheet carrier densities ($n_i$), and the conductance tensor is given by the summation of the individual components:

$$G_{xx} = \sum_i \frac{e|n_i|\mu_i}{1 + \mu_i^2 B^2}$$

$$G_{xy} = \sum_i \frac{e n_i \mu_i^2 B}{1 + \mu_i^2 B^2}$$

Here, $G_{xx}$ is the longitudinal conductance, $G_{xy}$ is the Hall conductance, $B$ is the magnetic field, and $e$ is the electron charge. Experimentally the resistances are the measured quantities, so $G$ can be inverted as

$$R_{xx} = \frac{G_{xx}}{G_{xx}^2 + G_{xy}^2},$$

and

$$R_{xy} = \frac{G_{xy}}{G_{xx}^2 + G_{xy}^2}.$$

The resistivities are then calculated as $\rho = R*thickness$.

For a two-band electron-hole model, as observed for Pd$_3$Bi$_2$Se$_2$, $n_i$ is taken to be less than zero for the electron-like pocket and greater than zero for the hole-like pocket. Further, to fit the experimental data these two conduction channels require four free parameters ($n_1,\mu_1,n_2,\mu_2$). However, this can be reduced to two parameters since $R_{xx}$ and $R_{xy}$ are well-defined functions of $n_i$ and $\mu_i$ in the zero-field limit[3]. Here $R_{xx}(B\rightarrow0) = f_{xx}(n_1,\mu_1,n_2,\mu_2)$ and $R_{xy}(B\rightarrow0) = f_{xy}(n_1,\mu_1,n_2,\mu_2)$. These two equations can then be solved for two of the variables (for example, $n_1$, and $n_2$) then substituted into the above expression. This results in an equation with two parameters (for example, $\mu_1$ and $\mu_2$). As shown in the main text, this gives an excellent fit to the experimental data. This shows that the Hall effect could be described by a two-band model that consisted of a hole pocket (carrier density ~2-4×10$^{22}$ cm$^{-3}$ and mobility ~100 cm$^2$V$^{-1}$s$^{-1}$) and an electron pocket (carrier density 1-3×10$^{19}$ cm$^{-3}$ and ~1000 cm$^2$V$^{-1}$s$^{-1}$).